\documentclass[runningheads]{llncs}
\usepackage{graphicx}
\usepackage{epsfig}
\usepackage{dsfont}
\usepackage{amsmath}
\usepackage{amssymb}
\usepackage{color}
\usepackage{boldline,multirow}
\usepackage[normalem]{ulem}
\useunder{\uline}{\ul}{}
\usepackage{siunitx}
\usepackage{booktabs}

\usepackage{cite}
\usepackage{cases}
\usepackage{hyperref}
\usepackage{bbding}
\usepackage{bm}
\begin{document}
\title{MBA-Net: SAM-driven Bidirectional Aggregation Network for Ovarian Tumor Segmentation}

\titlerunning{SAM-driven Bidirectional Aggregation Network}

\author{Yifan Gao\inst{1,2} \and Wei Xia\inst{2} \and Wenkui Wang\inst{3} \and Xin Gao\inst{2}\Envelope}

\authorrunning{Y. Gao et al.}

\institute{School of Biomedical Engineering (Suzhou), Division of Life Science and Medicine, University of Science and Technology of China, Hefei, China\\ \and
	Suzhou Institute of Biomedical Engineering and Technology, Chinese Academy of Sciences, Suzhou, China \and State Key Laboratory of Ultra-precision Machining Technology, Department of Industrial and Systems Engineering, The Hong Kong Polytechnic University, Kowloon, Hong Kong SAR, PR China \\ \email{yifangao@mail.ustc.edu.cn} \\
	\email{Wang.wenkui@polyu.edu.hk} \\
	\email{xiaw@sibet.ac.cn, xingaosam@163.com}
}
\maketitle
\begin{abstract}
	Accurate segmentation of ovarian tumors from medical images is crucial for early diagnosis, treatment planning, and patient management. However, the diverse morphological characteristics and heterogeneous appearances of ovarian tumors pose significant challenges to automated segmentation methods. In this paper, we propose MBA-Net, a novel architecture that integrates the powerful segmentation capabilities of the Segment Anything Model (SAM) with domain-specific knowledge for accurate and robust ovarian tumor segmentation. MBA-Net employs a hybrid encoder architecture, where the encoder consists of a prior branch, which inherits the SAM encoder to capture robust segmentation priors, and a domain branch, specifically designed to extract domain-specific features. The bidirectional flow of information between the two branches is facilitated by the robust feature injection network (RFIN) and the domain knowledge integration network (DKIN), enabling MBA-Net to leverage the complementary strengths of both branches. We extensively evaluate MBA-Net on the public multi-modality ovarian tumor ultrasound dataset and the in-house multi-site ovarian tumor MRI dataset. Our proposed method consistently outperforms state-of-the-art segmentation approaches. Moreover, MBA-Net demonstrates superior generalization capability across different imaging modalities and clinical sites.
	
\end{abstract}
\section{Introduction}
Ovarian cancer is one of the most lethal gynecological malignancies, accounting for over 200,000 deaths globally each year \cite{sung2021global}. Early and accurate detection and delineation of ovarian tumors is crucial for improving prognosis, treatment planning, and survival rates. However, the manual segmentation of ovarian tumors from medical images is a tedious, time-consuming, and subjective process, often leading to high inter- and intra-observer variability \cite{kiefer2018inter,schmidt2023probabilistic}. This has motivated the development of automated computational methods for ovarian tumor segmentation, which can provide objective and reproducible results, thereby assisting clinicians in making more informed decisions and ultimately improving patient care \cite{rudie2019emerging}. 

Traditional segmentation methods like graph cuts and level sets rely heavily on hand-crafted features and prior assumptions about tumor appearance, limiting their performance and generalization capability \cite{liu2014tumor,sawyer2019evaluation}. Recent years have seen a paradigm shift towards deep learning for medical image segmentation. Convolutional neural networks (CNN) and Transformers have achieved state-of-the-art performance by automatically learning hierarchical representations directly from data \cite{liu2022multi,hu2023uncertainty,chen2023improving,hu2023deep,gao2023anatomy}.

Nevertheless, automatic ovarian tumor segmentation has persisted as an open challenge, largely due to the significant diversity across ovarian cancer subtypes \cite{nougaret2019ovarian,wang2022computed}. Over 30 histological classifications have been documented, with tumors exhibiting substantial variance in morphology, growth patterns, and imaging phenotypes \cite{soslow2008histologic}. Irregular tumor shape and ambiguous tumor boundaries further complicate segmentation tasks, especially in ultrasound \cite{pham2023comprehensive}, which is heavily utilized for ovarian cancer screening and diagnosis. Additionally, the lack of data on some relatively rare subtypes has constrained model performance and generalizability across imaging modalities and clinical centers \cite{zhao2022multi}.

Recent advancements in visual foundation models, such as the Segment Anything Model (SAM) \cite{kirillov2023sam}, have demonstrated exceptional promise in imparting model generalization for medical image analysis \cite{gao2023desam}. SAM has exhibited the aptitude for few-shot generalization and shape sensitivity that could potentially address challenges in automatic ovarian tumor segmentation.

In this work, we present MBA-Net, a novel SAM-driven bidirectional aggregation network that effectively adapts and extends SAM's powerful segmentation capabilities to the ovarian tumor domain. The core innovation lies in synergistically composing SAM with a specialized CNN encoder branch through bidirectional feature aggregation. Specifically, our proposed network employs a hybrid encoder architecture consisting of two parallel branches: 1) a prior branch that inherits the SAM encoder to capture robust segmentation priors, and 2) a domain branch composed of multiple residual blocks, specifically designed for extracting domain-specific features. The shallow features from the prior branch are progressively aggregated into the domain branch via a cascade of robust feature injection networks (RFIN). RFIN enables the domain branch to incorporate the robust segmentation priors captured by the prior branch, guiding the extraction of domain-specific features. Conversely, the deep features from the domain branch are reciprocally integrated into the prior branch through several domain knowledge integration networks (DKIN). DKIN allows the prior branch to leverage the domain-specific knowledge extracted by the domain branch, enhancing its ability to adapt to the unique characteristics of ovarian tumors.

We extensively evaluate MBA-Net on two challenging datasets: a public dataset of ovarian tumor ultrasound images and an in-house dataset of ovarian tumor MRI scans. Comprehensive experiments and comparisons against state-of-the-art methods demonstrate MBA-Net's superior segmentation accuracy, robustness to tumor heterogeneity, and strong generalization across imaging modalities and clinical centers. 

\begin{figure}
	\includegraphics[width=0.9\textwidth]{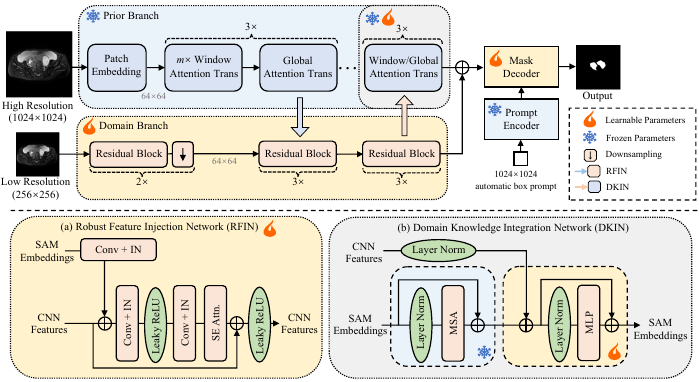}
	\centering
	\caption{The architecture of MBA-Net. It has two parallel branches: (1) the prior branch captures robust features, and (2) the domain branch extracts domain-specific features. The bidirectional feature aggregation between the two branches is achieved through (a) the robust feature injection network (RFIN) that injects SAM embeddings into the domain branch and (b) the domain knowledge integration network (DKIN) that integrates CNN features into the prior branch. The prompt encoder and mask decoder are inherited from SAM for mask prediction.} 
	\label{fig1}
\end{figure}

\section{Methodology}
\subsection{Overview of MBA-Net}
MBA-Net is a novel architecture that synergistically integrates the powerful general segmentation capabilities of the SAM with a domain-specific branch tailored for ovarian tumor segmentation. As depicted in Fig. \ref{fig1}, the encoder of MBA-Net comprises two parallel branches: the prior branch and the domain branch. The prior branch leverages the pre-trained image encoder from SAM. This branch operates on high-resolution images to capture global contextual information. In contrast, the domain branch is specifically designed to extract domain-specific features from lower-resolution images. It consists of a series of residual blocks with squeeze-and-excitation modules  \cite{hu2018squeeze}, enabling the learning of discriminative tumor representations.

The decoder of MBA-Net inherits the mask decoder from SAM. The fused features from the encoder are passed through the decoder to obtain the final segmentation output. Similarly, MBA-Net also inherits SAM's prompt encoder. We feed a box prompt the same size as the image into the prompt encoder for automatic segmentation.

During training, the prior and domain branches process their respective input images to extract image embeddings. These embeddings are then fused through the bidirectional aggregation, enabling the model to learn a unified representation that combines the strengths of both branches. The decoder takes the fused embeddings as input and generates the segmentation masks.

\subsection{The Domain Branch}
The domain branch is designed to extract domain-specific features from the input image $x_c$ with a resolution of 256×256. This branch consists of 8 layers, each implemented as a residual block with channel attention. 

Let $f_c^i$ denote the output feature map of the $i$-th layer in the domain branch. The first two layers of the domain branch are responsible for downsampling the input image $x_c$ from a resolution of 256×256 to 64×64. This is achieved through strided convolutions with a kernel size of 3×3 and a stride of 2. Starting from the third layer, the domain branch receives embeddings from the prior branch at specific locations. The prior branch, which operates on a higher-resolution image $x_s$, has a total of 4m layers arranged in a transformer architecture (m=8 in our work). The layers at positions \{m, 2m, 3m\} of the prior branch are global attention layers, enabling the capture of global contextual information. These global attention layers are connected to the \{3, 4, 5\} layers of the domain branch, respectively.

The feature maps are transmitted back to the prior branch in the later layers of the domain branch, specifically layers \{6, 7, 8\}. These layers aim to provide domain-specific features to the prior branch, enabling it to adapt its representations to the characteristics of ovarian tumors. The feature transmission is achieved through residual connections, where the output feature maps of the domain branch are added to the embeddings of the prior branch at layers \{4m-2, 4m-1, 4m\}. Finally, the output of the domain branch and the prior branch are combined through element-wise addition. We will introduce the details of bidirectional feature aggregation in the next section.

\subsection{Bidirectional Feature Aggregation}
Our architecture is fundamentally anchored in its bidirectional feature aggregation mechanism, a critical structure enabling seamless information flow between the prior and domain branches. This mechanism fosters a dynamic interchange of features and effectively enhances the network's ability.

\subsubsection{Robust Feature Injection Network:}
As illustrated in Fig. \ref{fig1}a, we devise lightweight residual connections to enable the robust features of the prior branch to be injected into the shallow layers of the domain branch.

We denote the output embedding generated by the i-th layer of the prior branch as $f^s_i\in\mathbb{R}^{N\times C}$, where $N$ is the number of tokens. Likewise, the j-th residual block generates a counterpart feature map $f^c_j\in \mathbb{R}^{H\times W \times C_c}$, where $H$, $W$, and $C_c$ represent its height, width and number of channels respectively.

To fuse robust features into the domain branch, a convolution layer is first applied to project $f^s_i$ into an aligned space:

\begin{equation}
	\Tilde{f}^s_i = \sigma(IN(W_{af}*R(f^s_i) + b_{af}))
\end{equation}

where $W_{af} \in \mathbb{R}^{C\times C_c\times k\times k}$ and $b_{af}\in \mathbb{R}^{C_c}$ represent the weight and bias of the convolution layer transforming $f^s_i$ from $C$ to $C_c$ channels. $R$ represents reshaping two-dimensional embeddings into three-dimensional features. IN and $\sigma$ denote instance normalization and Leaky ReLU activation. The projected embedding $\Tilde{f}^s_i$ is then combined with $f^c_j$ using residual addition:

\begin{equation}
	f^{c\prime}_j = f^c_j + \Tilde{f}^s_i, \quad \forall j \in \mathcal{J}
\end{equation}

where $\mathcal{J}$ denotes the index set \{3, 4, 5\} for low-level residual blocks.

\subsubsection{Domain Knowledge Integration Network:}
Our proposed DKIN is shown in Fig. \ref{fig1}b. We denote the output feature map of the j-th residual block as $f^c_j\in\mathbb{R}^{H\times W\times C}$ and the output embedding generated by the i-th transformer layer as $f^s_i\in\mathbb{R}^{N\times C}$.

To enable domain-specific features to enhance transformer layers in the prior branch, we apply residual addition to combine  $f^c_j$ with $f^s_i$:

\begin{equation}
	{f^s_i}' = \mathrm{MSA}(\mathrm{LN}(f^s_i)) + \mathrm{LN}(R(f^c_j)) + f^s_i, \quad \forall i \in \mathcal{I}
\end{equation}

where LN and MSA denote layer normalization and multi-head self attention module. $R$ represents reshaping three-dimensional features into two-dimensional embeddings. $\mathcal{I}$ denotes the index set \{4m-2, 4m-1, 4m\} for high-level transformer layers. This top-down dissemination enables advanced domain knowledge to directly guide deeper transformer layers in a tailored manner.

\section{Experiments and Results}

\subsection{Datasets and Implementation Details}
To validate the effectiveness of our proposed MBA-Net, we evaluate our method on a public multi-modal ovarian tumor ultrasound segmentation dataset \cite{zhao2022multi} and an in-house multi-site ovarian tumor MRI dataset. We employ a 6:1:3 split for training, validation, and testing for both datasets. Additionally, to assess the robustness of the model, we perform cross-domain and cross-modality testing using all available images.

After excluding samples of normal ovaries, the ultrasound dataset consists of seven different tumor types: chocolate cyst, serous cystadenoma, teratoma, theca cell tumor, simple cyst, mucinous cystadenoma, and high-grade serous carcinoma. We utilize the ultrasound images for training, evaluating the model's performance and assessing its robustness by contrast-enhanced ultrasonography images.

The MRI dataset comprises 493 patients from five medical centers, encompassing five distinct tumor types: serous cystadenocarcinoma, mucinous cystadenocarcinoma, serous cystadenoma, mucinous cystadenoma, and clear cell ovarian carcinoma. Ethical approval for this study was granted by the Jinshan Hospital of Fudan University's ethics committee (reference number 2018-30), and the requirement for informed consent was waived due to its retrospective nature. We employ data from one center for training and evaluating the model's performance while leveraging the other four centers to assess its robustness.

We compare MBA-Net with several state-of-the-art medical image segmentation networks, including U-Net \cite{ronneberger2015u}, TransFuse \cite{zhang2021transfuse}, TransUNet \cite{chen2021transunet}, and UTNet \cite{gao2021utnet}. We adopt the widely used Dice score as the evaluation metric to measure the performance of different methods. We apply consistent data augmentation techniques to ensure a fair comparison among different methods, including random intensity scaling, random intensity shifting, and random flipping. The training and testing processes are conducted on an RTX 4090 GPU with 24GB of memory. Each model is trained for 50 epochs. During the training phase, we employ the SGD optimizer with a momentum of 0.99 and a weight decay of 1e-4. The batch size is set to 2. The initial learning rate is 3e-4, and an exponential decay with a power of 0.9 is applied at the end of each training epoch.

\begin{table}[]
	\centering
	\caption{Quantitative comparison of our proposed MBA-Net against other state-of-the-art approaches on the multi-modality ovarian tumor ultrasound dataset. The best results are highlighted in bold.}
	\label{tab:2}
	\resizebox{\columnwidth}{!}{%
		\begin{tabular}{cccccccc}
			\toprule
			Methods   & Chocolate cyst           & \begin{tabular}[c]{@{}c@{}}Serous \\ cystadenoma\end{tabular}       & Teratoma                 & \begin{tabular}[c]{@{}c@{}}Theca cell \\ tumor\end{tabular}         & Simple cyst              & \begin{tabular}[c]{@{}c@{}}Mucinous \\ cystadenoma\end{tabular}  & \begin{tabular}[c]{@{}c@{}}High grade \\ serous\end{tabular}        \\ \midrule
			\multicolumn{8}{l}{\textbf{Dataset:   Multi-modality ovarian tumor ultrasound  \cite{zhao2022multi}}}                                                                                                                       \\ \midrule
			U-Net     & 86.48$\pm$15.88          & 92.40$\pm$9.83           & 83.45$\pm$16.99          & 89.32$\pm$6.15           & 89.67$\pm$13.83          & 89.10$\pm$11.68         & 83.16$\pm$12.54          \\
			TransFuse    & 82.40$\pm$21.09          & 90.52$\pm$10.69          & 80.15$\pm$15.30          & 89.19$\pm$7.07           & 90.64$\pm$12.26          & 90.41$\pm$8.52          & 82.34$\pm$11.02          \\
			TransUNet & 82.38$\pm$19.67          & 91.06$\pm$11.53          & 79.33$\pm$16.58          & 89.24$\pm$8.16           & 87.18$\pm$12.33          & 89.32$\pm$10.34         & 81.31$\pm$11.77          \\
			UTNet     & 88.54$\pm$13.80          & \textbf{93.78$\pm$6.42}  & 86.59$\pm$13.58          & 91.90$\pm$4.67           & 91.02$\pm$13.97          & 90.32$\pm$6.97          & 84.22$\pm$11.44          \\
			MBA-Net   & \textbf{90.39$\pm$11.01} & 92.98$\pm$5.58           & \textbf{89.34$\pm$7.75}  & \textbf{92.31$\pm$5.11}  & \textbf{92.22$\pm$11.92} & \textbf{90.67$\pm$6.70} & \textbf{88.75$\pm$7.16}  \\ \midrule
			\multicolumn{8}{l}{\textbf{Cross-modality   performance}}                                                                                                                                             \\ \midrule
			U-Net     & 69.44$\pm$27.50          & 83.75$\pm$13.10          & 60.43$\pm$29.05          & 77.75$\pm$19.59          & 75.47$\pm$30.43          & 85.98$\pm$13.60         & 48.85$\pm$32.38          \\
			TransFuse    & 76.74$\pm$17.96          & 83.52$\pm$13.77          & 62.57$\pm$28.01          & 76.30$\pm$18.81          & 75.42$\pm$31.42          & 88.61$\pm$7.50          & 47.55$\pm$33.59          \\
			TransUNet & 76.13$\pm$19.43          & 85.74$\pm$13.92          & 60.75$\pm$28.23          & 77.03$\pm$18.53          & 78.84$\pm$26.21          & 87.03$\pm$11.84         & 45.88$\pm$33.38          \\
			UTNet     & 75.55$\pm$14.82          & 81.53$\pm$14.18          & 61.28$\pm$30.26          & 73.77$\pm$19.07          & 77.51$\pm$25.54          & 81.56$\pm$7.26          & 50.34$\pm$31.61          \\
			MBA-Net   & \textbf{78.89$\pm$13.36} & \textbf{86.51$\pm$12.68} & \textbf{68.86$\pm$26.01} & \textbf{80.64$\pm$12.67} & \textbf{82.98$\pm$21.18} & \textbf{90.81$\pm$5.85} & \textbf{70.60$\pm$22.79} \\ \bottomrule
		\end{tabular}%
	}
\end{table}

\subsection{Results}
We evaluate the performance of MBA-Net and compare it with state-of-the-art segmentation methods on the multi-modality ovarian tumor ultrasound dataset. The results are summarized in Table \ref{tab:2}. MBA-Net achieves the highest average Dice score, outperforming all other methods. To assess the robustness of MBA-Net, we perform cross-modality evaluation by testing the model on contrast-enhanced ultrasonography images. As shown in Table \ref{tab:2}, MBA-Net exhibits superior generalization capability, surpassing all other methods by a significant margin. Remarkably, MBA-Net has substantially improved the segmentation of teratoma and high-grade serous carcinoma. These results highlight the effectiveness of incorporating SAM's robust features, which enables MBA-Net to better adapt to variations in imaging modalities and tumor characteristics.

\begin{table}[]
	\centering
	\caption{Quantitative comparison of our proposed MBA-Net against other state-of-the-art approaches on the multi-site ovarian tumor MRI dataset. The best results are highlighted in bold.}
	\label{tab:3}
	\resizebox{\columnwidth}{!}{%
		\begin{tabular}{cccccc}
			\toprule
			Methods   & \begin{tabular}[c]{@{}c@{}}Serous \\ cystadenocarcinoma\end{tabular} & \begin{tabular}[c]{@{}c@{}}Mucinous \\ cystadenocarcinoma\end{tabular} & \begin{tabular}[c]{@{}c@{}}Serous \\ cystadenoma\end{tabular} & \begin{tabular}[c]{@{}c@{}}Mucinous \\ cystadenoma\end{tabular} & \begin{tabular}[c]{@{}c@{}}Clear cell \\ ovarian carcinoma\end{tabular} \\ \midrule
			\multicolumn{6}{l}{\textbf{Dataset: Multi-site   ovarian tumor MRI}}                                                                                                                                                                                                                                                                                                  \\ \midrule
			U-Net     & 78.33$\pm$21.94                                                      & 82.09$\pm$20.75                                                        & 82.24$\pm$15.22                                               & 90.02$\pm$9.80                                                  & 89.47$\pm$10.10                                                         \\
			TransFuse & 81.17$\pm$19.27                                                      & 81.48$\pm$17.97                                                        & 79.44$\pm$18.06                                               & 92.94$\pm$7.59                                                  & 89.88$\pm$12.23                                                         \\
			TransUNet & 80.91$\pm$18.41                                                      & 83.56$\pm$15.72                                                        & 85.16$\pm$12.53                                               & 91.36$\pm$8.62                                                  & 86.92$\pm$10.08                                                          \\
			UTNet     & 79.69$\pm$21.05                                                      & 83.82$\pm$16.67                                                        & \textbf{86.69$\pm$14.40}                                      & 91.45$\pm$6.83                                                  & 91.61$\pm$8.94                                                          \\
			MBA-Net   & \textbf{83.51$\pm$17.14}                                             & \textbf{86.75$\pm$12.18}                                               & 86.42$\pm$11.96                                               & \textbf{93.79$\pm$6.55}                                         & \textbf{94.73$\pm$5.50}                                                 \\ \midrule
			\multicolumn{6}{l}{\textbf{Cross-site   performance}}                                                                                                                                                                                                                                                                                                                 \\ \midrule
			U-Net     & 74.05$\pm$23.88                                                      & 69.75$\pm$23.68                                                        & 71.14$\pm$25.39                                               & 90.33$\pm$11.90                                                 & 81.73$\pm$19.96                                                         \\
			TransFuse & 73.11$\pm$22.06                                                      & 70.79$\pm$21.41                                                        & 75.72$\pm$20.26                                               & 88.49$\pm$14.74                                                 & 80.29$\pm$21.33                                                         \\
			TransUNet & 72.99$\pm$20.11                                                      & 75.42$\pm$17.05                                                        & 78.72$\pm$18.88                                               & 89.54$\pm$15.44                                                 & 85.84$\pm$14.62                                                         \\
			UTNet     & 75.79$\pm$22.41                                                      & 77.92$\pm$18.13                                                        & 74.40$\pm$21.56                                               & 90.77$\pm$13.82                                                 & 84.47$\pm$13.77                                                         \\
			MBA-Net   & \textbf{80.35$\pm$19.29}                                             & \textbf{83.47$\pm$16.26}                                               & \textbf{83.18$\pm$14.97}                                      & \textbf{91.19$\pm$11.76}                                        & \textbf{89.11$\pm$11.92}                                                \\ \bottomrule
		\end{tabular}%
	}
\end{table}

We further validate the performance of MBA-Net on the multi-site ovarian tumor MRI dataset. The results are presented in Table \ref{tab:3}. MBA-Net consistently outperforms other methods across most tumor types. MBA-Net also demonstrates superior generalization ability in the cross-site setting. These results showcase the effectiveness of MBA-Net in handling variations across different clinical sites and scanners.

\begin{figure}
	\includegraphics[width=0.68\textwidth]{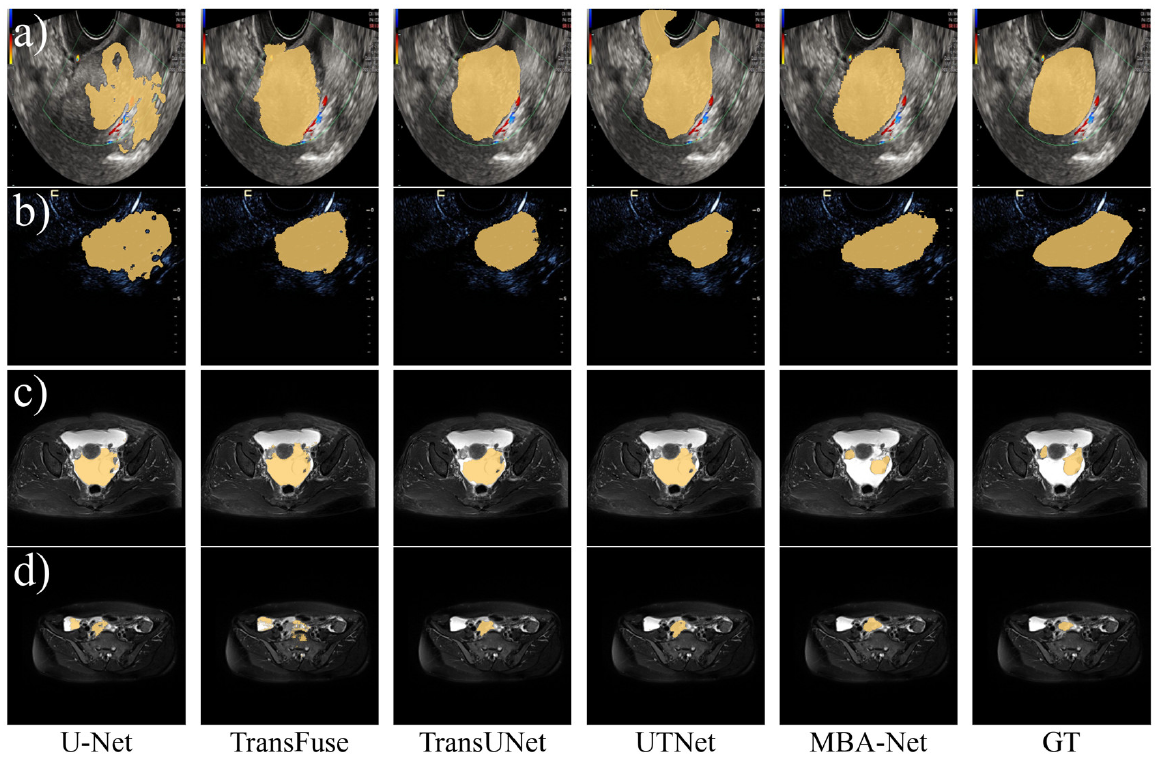}
	\centering
	\caption{Visual comparison of segmentation results obtained by different methods on representative examples from (a-b) the multi-modality ovarian tumor ultrasound dataset and (c-d) the multi-site ovarian tumor MRI dataset. GT: ground truth.} 
	\label{fig2}
\end{figure}

Fig. \ref{fig2} presents the visual comparison of the segmentation results obtained by different methods on representative examples from both the multi-modality ovarian tumor ultrasound dataset and the multi-site ovarian tumor MRI dataset, illustrating the superior performance of MBA-Net in accurately delineating ovarian tumors with diverse morphological characteristics and imaging modalities.

\subsection{Ablation Studies}
To investigate the effectiveness of the proposed bidirectional feature aggregation mechanism, we conduct ablation studies on the ovarian tumor ultrasound dataset. We evaluate the impact of varying the number of RFIN and DKIN modules on the segmentation performance. The results are summarized in Table \ref{tab:4}. The best performance is achieved when both RFIN and DKIN modules are utilized, with the default setting of 3 RFIN and 3 DKIN modules yielding an average Dice score of 90.75\%. This result validates the effectiveness of the bidirectional feature aggregation mechanism in leveraging the complementary strengths of the prior branch and the domain branch.

\begin{table}[]
	\centering
	\caption{Ablation studies of the different feature fusion strategies.}
	\label{tab:4}
	\resizebox{0.8\columnwidth}{!}{%
		\begin{tabular}{ccc|ccc|ccc}
			\toprule
			RFIN & DKIN & Avg. Dice (\%) & RFIN &  DKIN & Avg. Dice (\%) & RFIN & DKIN & Avg. Dice (\%) \\ \midrule
			0             & 1           & 85.84            & 0           & 3           & 86.15            & 0           & 6           & 86.58            \\
			1             & 1           & 88.69            & 1           & 3           & 88.43            & 1           & 6           & 89.06            \\
			2             & 1           & 88.01            & 2           & 3           & 90.19            & 2           & 6           & 89.82            \\
			3             & 1           & 88.25            & 3           & 3           & \textbf{90.75}   & 3           & 6           & 90.47            \\ \bottomrule
		\end{tabular}%
	}
\end{table}

\section{Conclusion}
In this paper, we present MBA-Net, a novel architecture for ovarian tumor segmentation that integrates the powerful segmentation capabilities of the SAM with domain-specific knowledge through bidirectional feature aggregation. The proposed network employs a hybrid encoder architecture, where the encoder consists of a prior branch, which inherits the SAM encoder to capture robust segmentation priors, and a domain branch, which is specifically designed to extract domain-specific features. The robust feature injection network (RFIN) and the domain knowledge integration network (DKIN) facilitate the bidirectional flow of information between the two branches, enabling MBA-Net to leverage the complementary strengths of both branches. Extensive experiments on multi-modality ovarian tumor ultrasound and multi-site ovarian tumor MRI datasets demonstrate the superiority of MBA-Net in terms of segmentation accuracy, robustness to tumor heterogeneity, and generalization capability across different imaging modalities and clinical sites.

\begin{credits}
	\subsubsection{\ackname} This work was supported in part by National Science Foundation of China under Grant 82372052, in part by Taishan Industrial Experts Program under Grant tscx202312131, in part by Key Research and Development Program of Shandong under Grant 2021SFGC0104, in part by Science Foundation of Shandong under Grant ZR2022QF071 and ZR2022QF099, and in part by Key Research and Development Program of Jiangsu under Grant BE2021663 and BE2023714.
	
	\subsubsection{\discintname}
	The authors have no competing interests to declare that are relevant to the content of this article.
\end{credits}

\bibliographystyle{splncs04}
\bibliography{mybibliography}
	
\end{document}